\documentclass[sigconf]{acmart}

\usepackage{graphicx}
\usepackage{subfigure}
\usepackage{multirow}
\usepackage{calc}

\usepackage{textcomp}
\usepackage{siunitx}
\usepackage{soul,xcolor}
\usepackage{gensymb}

\AtBeginDocument{%
  \providecommand\BibTeX{{%
    \normalfont B\kern-0.5em{\scshape i\kern-0.25em b}\kern-0.8em\TeX}}}

\setcopyright{acmcopyright}
\copyrightyear{2018}
\acmYear{2018}
\acmDOI{XXXXXXX.XXXXXXX}

\acmConference[AHs '23]{Augmented Humans (Assistive Augmentation Workshop)}{March 12--14, 2023}{Glasgow, UK}
\acmPrice{15.00}
\acmISBN{978-1-4503-XXXX-X/18/06}

\begin{document}


\title{Designing for Affective Augmentation: \\ Assistive, Harmful, or Unfamiliar?}
%

\author{Abdallah El Ali}
\email{aea@cwi.nl}
\orcid{0000-0002-9954-4088}
\affiliation{%
  \institution{Centrum Wiskunde \& Informatica (CWI)}
  \city{Amsterdam}
  \country{The Netherlands}
}


%

\renewcommand{\shortauthors}{El Ali}

\begin{abstract}



In what capacity are affective augmentations helpful to humans, and what risks (if any) do they pose? In this position paper, we outline three works on affective augmentation systems, where our studies suggest these systems have the ability to influence our cognitive, affective, and (social) bodily perceptions in perhaps unusual ways. We provide considerations on whether these systems, outside clinical settings, are assistive, harmful, or as of now largely unfamiliar to users.


%



\end{abstract}

\begin{teaserfigure}
        \centering
        \setlength{\abovecaptionskip}{2pt}
        \includegraphics[width=0.8\textwidth]{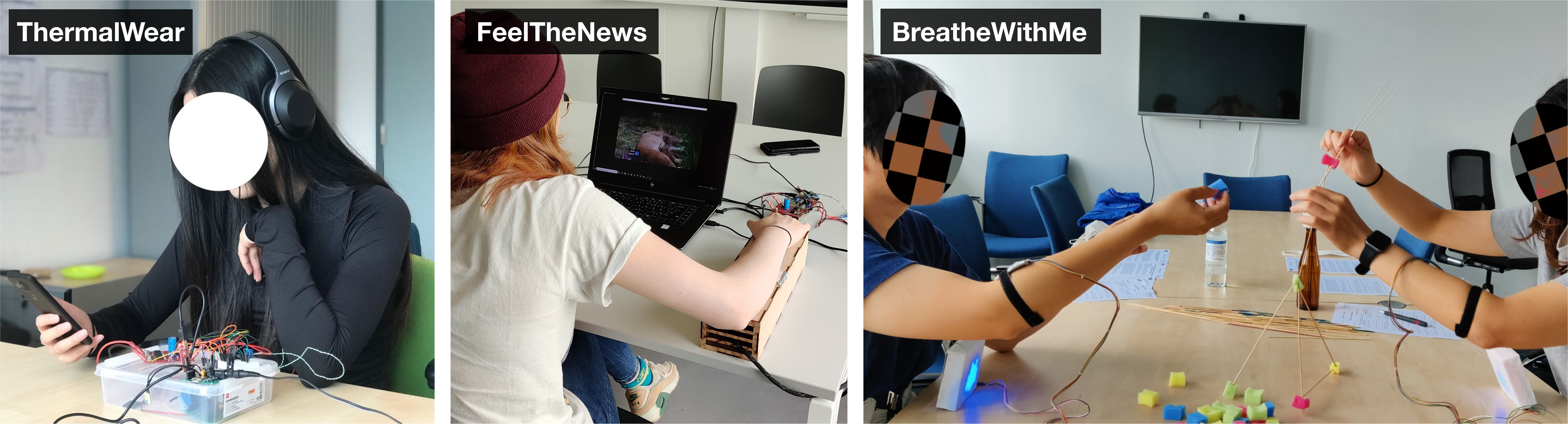}
                \caption{Left: ThermalWear \cite{Elali2020}. Middle: FeelTheNews \cite{Ooms2023}. Right: BreatheWithMe \cite{Elali2023}.}                
        \label{fig:teaser}
        \Description{Left: ThermalWear. Middle: FeelTheNews. Right: BreatheWithMe}
    \end{teaserfigure}



\maketitle

\section{Introduction}

We are entering a digital wave where the user experience is transforming through immersive, interactive, and multi-sensory technologies, where our emotion and affective states play a strong role. Human emotion and affect can be seen as mental and physical states that touch every aspect of human life throughout every (waking) moment, and are an integral part of human cognition, behavior, and social interaction. Within the Distributed \& Interactive Systems research group, we focus on designing and developing \textbf{Affective Interactive Systems}, where one key area is on Affective Augmentation: How can we develop systems that can augment our physical / virtual bodies and sensory perception to enhance, modify, or diminish our affective states and (social) interactions?

In this position paper, we discuss so-called \textbf{Affective Augmentation Systems}. Advancements in biosensing and actuation enable us to not only visualize and share hidden physiological data, but also to create artificial haptic sensations \cite{Schoeller2019} (e.g., thermal, vibrotactile), which together can potentially enhance our mind, body, and social connections (cf., \cite{Feijt2021}). This can create unfamiliar human-human and human-machine affective interactions, where the line between betterment (assistive, empowering, enabling, augmenting) and harm \textit{may} blur. Given the private nature of our emotions and internal bodily states that go well beyond overt motor actions, we suspect this may be a thin line to tread, and potentially impede the daily adoption and acceptability of such systems outside clinical contexts without responsible design considerations. Below we give a brief overview of three research prototypes to help frame the discussion.

\section{Research Prototypes}

\subsection{ThermalWear}

With ThermalWear (Fig.\ref{fig:teaser}, left) \cite{Elali2020}, a wearable on-chest thermal display, we aimed to address how thermal stimulation can support emotional prosody production, when this can be situationally or medically impaired. Since thermal displays have been shown to evoke emotions, we tested whether thermal stimulation can augment affective perceptions of neutrally-spoken smartphone voice messages. To that end, we synthesized and validated 12 neutrally-spoken voice messages, then tested in a within-subjects study (N=12) if thermal stimuli can augment their perception with affect. We found warm and cool stimuli generally increase arousal of voice messages, and increase / decrease message valence, respectively. Given these findings, ThermalWear provides opportunities that can for example support individuals with Autism Spectrum Disorder (ASD). Indeed prior work has shown impairment in emotional prosody processing in individuals with ASD \cite{Rosenblau2016}, and that this has a neural basis which results in increased reliance in cognitive control, attentional management, and reading of intentions \cite{Inge-Marie2012}. This lends credence to the potentially \textbf{assistive} use of thermally-augmented voice assistants, to support individuals with emotional prosody impairments.

\subsection{FeelTheNews}

With FeelTheNews (Fig.\ref{fig:teaser}, middle) \cite{Ooms2023}, we designed a prototype that combines vibrotactile and thermal stimulation (Audio-based Mapping, 70Hz/20\degree~C, 200Hz/40\degree~C) during news video watching. Emotion plays a key role in the emerging wave of immersive, multi-sensory audience news engagement experiences \cite{Pavlik2019,Goutier2021}. Since emotions can be triggered by somatosensory feedback (e.g., vibrotactile stimuli can facilitate conveying emotional meaning \cite{Ju2021}), in this work we explored in a within-subjects (N=20) study how augmenting news video watching with haptics can influence affective perceptions of news. We found that news valence and emotion intensity ratings were not affected by haptic stimulation, and that no stimulation was more comfortable than including stimulation. We also found that attention and engagement with the news can override haptic sensations, and crucially, accounting for users' perceived agency over their own emotional reactions is critical to avoid distrust. For the latter point, it brought to question users' sense of agency \cite{Cornelio2022,Mueller2020} should they experience any signal of \textit{``emotional hijacking"}. This requires understanding and minimizing any potentially adverse impacts of steering human emotion through haptic stimulation. As highlighted in prior work, the adoption and acceptance of such augmentations may require so-called sensory transparency \cite{Mueller2020}, where we would need to ensure responsible human-machine integration as we cross into the age of immersive and multi-sensory news media \cite{Goutier2021}.

\subsection{BreatheWithMe}

WithBreatheWithMe (Fig.\ref{fig:teaser}, right) \cite{Elali2023}, we designed a prototype that allows real-time sharing and receiving of breathing signals through visual, vibrotactile, or visual-vibrotactile modalities. While prior work showed that sharing breathing signals can provide insights into hidden experiences and enhance interpersonal communication \cite{macnaughton2020making}, it remains unclear how the modality of breath signals (visual, haptic) is socially interpreted during collaborative tasks. We ran a within-subjects study (15 pairs) to investigate the effects of modality on breathing synchrony, social presence, and overall user experience. We found no effects on breathing synchrony, and found the visual modality was preferred over vibrotactile feedback, despite no differences across social presence dimensions. Most relevant here, we found that BreatheWithMe was perceived to be an insightful window into others, however created \textbf{unfamiliar} experiences. These included gaining insight into others, social self-regulation, but also concerns over public data exposure (e.g., with the visual display) and social acceptability. As in FeelTheNews, it brought to question users' sense of bodily agency \cite{Cornelio2022} when continuously sharing breathing data with one another, where ambiguity, mis-inference, and manipulation can readily occur without a transparent control mechanism for \textit{when} it is appropriate for such signals to be represented and shown. For example, some participants sometimes deliberately manipulated their breathing to improve their self-image, whereas others used the other person's breathing as a social cue to self-regulate their own breathing.

\section{Affective Augmentation in Everyday Life: Cause for Concern?}


If we map the preceding three prototypes onto the design space of Assistive Augmentation \cite{Huber2018}, where on one axis is the degree of physical or cognitive disability, and on another axis the degree of bodily integration, we find the following: ThermalWear can be an assistive prototype, FeelTheNews an enabling prototype, and BreatheWithMe an augmentation prototype. From this lens, all seems fine -- these prototypes were created with the intention to improve our lives (whether by addressing a disability, enhancing our senses, or improving social interaction), in line with efforts toward sensible \cite{Mueller2020,Cornelio2022} and desirable \cite{Inami2022} human-computer integration. Recent work has touched upon emotion enhancement through interoceptive and emotion augmentation technologies \cite{Schoeller2022}. These include somatosensory interfaces and emotion prosthesis through creating artificial sensations, or interoceptive illusions for manipulating users' contextual cues to induce predictable drifts in their body perception \cite{Schoeller2019,Schoeller2022}, which may have beneficial clinical implications. 

However, based on the foregoing studies, something does not "feel" right with affective and emotional augmentations. In FeelTheNews and BreatheWithMe, for some participants, their sense of agency was threatened -- not at the conscious level of executing motor actions (e.g., \cite{Cornelio2022}), but at an affective perception level (which may have been little more than an afterthought were it not for deliberate reflectance during interviews). Another interpretation is that such prototypes are simply too early and require much wider population testing (cf., \cite{Schoeller2019}), where we are currently grappling with their \textit{unfamiliarity} and use cases outside of (clinical) affective neuroscience \cite{Schoeller2022}. On the other hand, some participants remarked FeelTheNews can help neutralize negative emotion if they can control the stimulation. Similarly with BreatheWithMe, some found the system to be a therapeutic social self-regulation tool. While these may be early works, they strongly hint that ensuring a good user experience would impact not just everyday social acceptability and concern \cite{Koelle2020,Kristen2011}, but uptake of such integration at a societal level \cite{Mueller2020}. This resonates with similar efforts across research areas (e.g., Hybrid Intelligence \cite{Akata2020}, Augmented Reality \cite{Liao2012}). While we currently do take the position that such systems are as of now unfamiliar, for this workshop, we would like to further discuss mechanisms, safeguards, and/or ethical guidelines for designing responsible affective augmentations from the start, touching upon questions such as: \textit{``If there is potential for harm from affective augmentations, does this necessitate a new design space?"}, \textit{``Who or what is accountable if my affective augmentation device twists my emotions and makes me harm myself or others?"}, or \textit{``Is this person (real or virtual) I'm interacting with using their augmentation device to deceive me by masking / modifying their affective signals?"}.

\section{Author Biography}

\textbf{Abdallah El Ali} is a research scientist in Human Computer Interactions (HCI) at  \href{https://www.cwi.nl/}{Centrum Wiskunde \& Informatica (CWI)} in Amsterdam within the  \href{https://www.dis.cwi.nl/}{Distributed \& Interactive Systems (DIS)} group. He is currently leading the research area on Affective Interactive Systems, with a focus on ground truth label acquisition techniques, emotion understanding and recognition, and affective augmentation systems that leverage physiological signals and multimodal feedback. He is also on the executive board for \href{https://chinederland.nl/}{CHI Nederland (CHI NL)}, an ACM SIGCHI Chapter that connects, supports, and represents the HCI community in the Netherlands. Website: \url{https://abdoelali.com}.

\section{Acknowledgements}

Thanks to all co-authors of the presented works, and the support of the Distributed \& Interactive Systems group at CWI.

\bibliographystyle{ACM-Reference-Format}
\bibliography{ah-ws-2023}


\end{document}